# An Engineer's Nightmare:
# 102 Years of Critical Robotics

Christopher Csíkszentmihályi
Information Science
Cornell University
Ithaca, NY
chris.csik@cornell.edu

*Abstract*— **A critical and re-configured HRI might look to the arts, where another history of robots has been unfolding since the Czech artist Karel Capek's critical robotic labor parable of 1921, in which the word robot was coined in its modern usage. This paper explores several vectors by which artist-created robots, both physical and imaginary, have offered pronounced contrasts to robots-as-usual, and offers directions as to how these more emancipated cousins might be useful to the field of HRI.**

## I. INTRODUCTION

Recent calls for an emerging paradigm in HRI, "critical robotics," [1] while both necessary and compelling, have elided the sustained and prolific contributions that creative artists have made to critical robotics for over 102 years [2]–[5]. These artistic robots have many of the qualities that might contribute to a new paradigm in HRI. For example, these art robots are grounded in rich historical and social contexts. Robotic art projects are conducted in public, as a means to reflect upon and engender a public debate about the values we have either wrongly embedded, or neglected to embed, in robots. Finally, they question and complicate the notion of utility so deeply engrained in research projects, expanding the possibility for different aspects of being and interaction, as utility unnecessarily constrains robotic possibilities.

While many of these artistic projects [e.g. by Rebecca Horn, Nam Jun Paik, Simon Penny] may not on first glance appear to offer enough bibliographic relevance to legitimately contribute to the rapidly maturing field of HRI, the degree to which they explicitly address sociocultural contexts – out of the lab and with members of the public – merits consideration by the HRI community. Similarly, while art robots might strike some researchers as too technically simple to be classified a robots, the degree to which much cutting-edge work in HRI is generated through "Wizard of Oz" or teleoperated experiments should prime its researchers to recognize, and benefit from, the narrative and symbolic contributions described in this paper.

Art robots and their interventions have changed and responded to their historical contexts over time, and when viewed in their context, have raised important and potentially generative questions for the wider domain of robotics. For example, why and how were creative artists able to: escape techno-positivist paradigms of function and automation? complicate behaviorist reductions of human psychology or essentialist assumptions about human nature? offer more diverse approaches to imagining robot gender and subjectivity? or reject the constant cross-pollination between the field of robotics and the enterprise of war? all while their contemporaries from science and engineering largely did not? Indeed, critical robotic experiments in the fine arts were often developed as self-conscious alternatives to contemporary technical research, and they provide a century of proof that radical reconfigurations of the field were both obvious and urgent enough to motivate artists to intervene, even though this often meant forgoing more lucrative traditional media.

Many art robot projects operate through deceptively simple juxtapositions, for example by providing robots with unexpected roles or contexts, such as Laura Kikauka and Norman White's (1988) *them fuckin' robots*, or by incorporating unexpected robotic components, e.g. Garnett Hertz's (2004) robot, governed by an embedded cockroach rather than a microcontroller. While these relatively straightforward switches of material or context may appear methodologically shallow, the public presentations of these robots often led to complex social interactions that surprised even the roboticist. [6] Loosening the constraints of function and materials thusly, art robots have demonstrated the opportunity to broaden the robotic design space in much the way speculative and critical design has done for HCI, while simultaneously casting light on the often-regressive assumptions of what we might call "uncritical robotics."

This paper offers a primer for HRI researchers to the critical approaches developed by roboticists in the fine arts—a set of reliable methods that have been tested and refined through public interactions with millions of human subjects. It will begin by looking at early contributions from works in the robotic art canon, then move to three areas where artists have configured radically different possibilities for robots: socioculturally complex behavior, emancipation from utility, and robot mortality.

## II. WHAT MAKES AN ART ROBOT?

Contemporary separations between art, engineering, and science date back only to the 1800s, and many early robots might rightly be claimed as belonging to the history of any of these contemporary endeavors. [7]–[11] For the sake of this paper, conceptual art robots will be defined as those coming after Capek's coinage, near the beginning of the European modern art movement. Prior to the 1950s, a rich history of visual machine art by Francis Picabia, Hannah Höch, Rube Goldberg, Ferdinand Léger, and others explored the utility of machines, human-machine relations and hybrids, and mechanical war through paintings, prints, sculptures, and performances. In parallel, artists like Laszlo Moholy-Nagy and Noam Gabo introduced motion into sculpture, and Marcel Duchamp played with ideas of mechanical gender and social order, but we cannot easily discern contemporary robots in these largely symbolic projects.

Jean Tinguely's sculptures from the early 1950s clearly transition toward robotic art. His *metamatic* automated painting machines, cobbled from electric motors and flexible metal rods, were able to sketch infinitely variable paintings and drawings through stochastic mechanical oscillations. They painted and sketched, and the images they put to paper were art, but the machines themselves were the "meta" aesthetic objects. The implication that Tinguely could leverage automation to make his abstract expressionist painting colleagues redundant was more the point of the *metamatics* than any of the images they generated. Also of note were his later performances, like *Homage a New York* [1960], a massive and polyphonous mechanism that represented the dynamism and intricacy of a city, and that ultimately destroyed itself as the Fire Department of the City of New York shut it down. Tinguely's works were largely assembled from older, often discarded technologies. We know that Tinguely could have used new materials, as he did in fact do so for many of his structural elements, but he chose old materials because of their cultural valances. A cow bell, painted gear, or the worn leg of a farmhouse chair: these materials were signifiers with terroir that tied his new creation to recognizable and rich histories. Tinguely exploited the randomness of loose springs and sloppy joints to engender complex motions and unpredictability, bequeathing his robots behavior and liveliness. The effect of quivering, unpredictable, intermittent motion gave his assemblages animate and human qualities long before computer scientists dubbed this "Perlin Noise."

Nam Jun Paik built what might be called the first humanoid art robot, *K-456*, in 1964. The robot was hermaphroditic, with both breasts and a penis; political, with a voice box made of John F. Kennedy recordings; and profane, and crude, as it shat beans. Cobbled from new and found materials, the robot was far from mechanically perfect: Paik's own brother, an engineer, asked if he could help to make the robot better. Paik's nephew recalled that "Nam said no, he liked it like that. He didn't want a more perfect robot." [12] Leaflets promoting his robot opera [1964] described the robot walking in Washington Square, Harlem, and "every streets and squares in New York." [5] Thousands saw and interacted with the robot through the late sixties, on the block and in art venues. After a decade in mothballs, the robot was revived in 1982 at the Whitney Museum where "Paik choreographed a performance titled *First Accident of the Twenty-First Century*, in which the robot was the victim of a car accident" on 75th and Madison Avenue: an inverted foreshadowing of contemporary Tesla manslaughter. [12]

Rebecca Horn's cybernetic mechanisms are similarly enculturated, from flying old suitcases and pianos (*Concert for Anarchy*, 1990) to massive architectural weeping breasts (*El Rio de la Luna*, 1992). [4] Scaled from tiny clockworks to warehouse-filling installations, each piece simultaneously fulfills relatively simple mechanical functions, but also complex and indeterminate sociocultural ones. "The works, many of which mechanically mimic intimate human gestures… caress, dance, explore, grope, flutter, falter, hesitate, shudder, stroke, tickle, whisper, and waver; all tender articulations of the deepest preoccupations of human subjectivity." [4]

The proto-robots of Horn, Paik, and Tinguely demonstrate three key elements that have come to define robotic art. First, they exhibit complex and subtle behaviors that are grounded in social, political, and cultural histories. Second, these behaviors are largely *emancipated* from utility, or serve ambiguous [13] utility. Third, they escape narratives of newness or innovation, embracing breakage, wear, pathos, and history. Finally, in addition to these three elements, they are empirical in the broad sense, in that they were created in tight feedback with the reactions of a great number of viewers in public settings. We now look at each of these elements in further detail.

### III. SOCIOCULTURALLY COMPLEX BEHAVIOR

While the machine art from the mid-twentieth century described above was grounded in kinetics – adding motion to painting or sculpture – by the 1980s many artists saw computation as a way to move beyond motion to behavior. In nearly all cases, the behavior was seen as social behavior, in relationship with a public. Sara Roberts' *Early Programming* (1988), albeit more a software agent than robot, puts the viewer in the position of a child dealing with a complex and controlling mother. [14] In contrast, the schematic metal robot legs in Heidi Kumao's works *Resist and Protest* [15] are immediately recognizable to any parent of a truculent six year old, as they stamp and kick in proportion to a viewer's proximity.

Artist France Cadet explains her motivation to build behavior through robots: "[I] wanted to have more interactive work… and to involve the audience because for me the audience is at least as important as the artwork." [16] Cadet hacked robot dog toys to change their behavior into a form of commentary on biopower and biotechnologies: "the robot is a medium to talk about animal and human relationships." Among her menagerie of robot dogs, the most compelling to audiences was one painted like a cow, which would happily graze until it developed Bovine Spongiform Encephalopathy (Mad Cow Disease) and collapsed twitching. Many viewers found this unsettling. In Cadet's formulation, it is the dialog with audiences that is the point, and the robots are decentered to become a medium of communication.

While artists making robots observe their audience, learning from these interactions to inform revisions or future designs, "user study" is not the main point of publicly exhibiting robot art. Rather, the main point is the experience generated *for* the viewer. Viewers are observed, their feedback is incorporated into revisions of the robot. But viewers are configured not as experimental subjects used to inform the field, but as *transforming subjects* who might be moved, changed, or informed by interactions with a socioculturally rich robot. I would posit that this is considerably different from how most HRI experiments work. Art viewers are autonomous, and can come or go from an exhibition; they may even pay for entry, rather than be paid by a researcher! Great care goes into designing a robot art project that will create a transformative moment for the public. Indeed, for an artist to have a viewer return to spend more time with a robot's behavior is considered a sign of a design's success.

One example of a compelling robot to which viewers often returned is *Petit Mal*, created by Simon Penny in the early 1990s, then a professor at Carnegie Mellon University. Penny sought to design an interaction "in the space of the body, in which kinesthetic intelligences, rather than 'literary-imagistic' intelligences play a major part." [17] Perhaps the only robot yet modeled on a form of epilepsy, *Petit Mal* loses motor control regularly thanks to its double-pendular body. To create a double pendulum robot that could be used safely in public was no small feat, and the odd amalgam of bicycle parts, ultrasonic and pyrosensitive sensors, and shelf-liner paper was, in Penny's words, "an engineer's nightmare." He sought a robot "which is truly autonomous; which is nimble and has 'charm'; that senses and explores architectural space and that pursues and reacts to people; that gives the impression of intelligence and has behavior which is neither anthropomorphic nor zoomorphic, but which is unique to its physical and electronic nature." Penny described *Petit Mal* as "an anti-robot," designed to avoid what he saw as the software-hardware (Cartesian mind-body) split typical in research robotics. "Hardware and software were considered as a seamless continuity, its behavior arises from the dynamics of its 'body'." Penny's rejection of Cartesianism was explicitly in reference to feminist critiques of the split, and directed at negating the extreme anti-body (and to be honest: body-hating) perspectives of AI founders like Marvin Minsky or Hans Moravec.

## IV. Emancipation from Utility

The robots described in the last section were designed to express behaviors, but not *subjugate behavior*. They interacted with viewers, but were not designed to serve the viewers. In my own experience with *Petit Mal*, the adjective that came to mind was "self-possessed." Viewers did not see it as a helper, but more as a mutually curious investigator, or an awkward dance partner. I'm not familiar enough with HRI to know if it has a word for this sort of self-possession, or the relationships it affords.

In contrast to self-possession, research robots generally manifest obsequiousness: design for utility. HRI experiments and their robots are built to discover new principles, but creating things for utility is a specific winnowing of possibilities. An example outside of robotics might be household pets. A wolf or ocelot does not make a good pet. Dogs and cats are domesticated and exhibit domesticated behaviors. But they retain other behaviors, and pet owners often find that the most fascinating aspects of their cats or dogs are the residual wild bits. How cat-like would cats be if they had been designed de novo for domesticity? Who would want a dog as straightforward as an Aibo? What, then, are roboticists omitting by nearly always assuming that robots will serve?

Cultural critics have long argued that the congenital defect of dominant American (or Silicon Valley) technology is its predication on, and implicit nostalgia for, the history of slavery. [18] Gregory Jerome Hampton, who analyzes narratives about robots, sees many parallels between contemporary robotic archetypes and those of black women during slavery identified by West. This includes not only physical labor robots, but also sex robots (Jezebels) and, crucially, *care-oriented robots* (Mammys). Hampton explains that the similarities are more literal than we might imagine: "To be clear, the archetypes and the actual female slaves being referred to in the antebellum period were not initially imagined as human 'women.' They were chattel or more appropriately, biological machines with crude gender assignments." [19] [20]

Hampton sees slavery embedded in science's and engineering's interests in robotics. This might appear to be an easily dismissed concern, because robots don't have consciousness, so they cannot be degraded by servitude. But historians have long observed that slavery warps and degrades not just slaves, but everyone in a slaving society. To Hampton, no one is immune to the inevitably corrosive nature of utilitarian robots: "The process of systematically or institutionally dehumanizing anything that is distinctly human-like is necessarily detrimental to all parties involved… we must make great efforts to be honest with ourselves about our desires and our history with slave/master relationships." [ibid, p. 81] US-based HRI researchers should indeed introspect on their own historical relationship to involuntary labor. Doing so would allow them to productively reconfigure many aspects of robotics. Chattel slavery was, however, not only a US problem. Roboticists based in University College London (London was long the epicenter of the slave trade, and the trade's profits endowed many of its most famous buildings and peers), University of Oslo and Aarhus University (Denmark-Norway profited from slavery until 1848), and Técnico Lisboa (Portugal was the first and last transatlantic slaver, from the 1400s - 1900s), among others, need to confront their historical complicity. Meanwhile HRI researchers from non-slaving cultures should carefully inspect their imports.

Anthropologist Jennifer Robinson argues that servitude is embedded in many gendered robots. Her analysis of robotics research is predicated on third wave feminists like Donna Haraway and Judith Butler, recognizing gender as a socio-material configuration. "In humans, gender is both a concept and performance embodied by females and males, a corporeal technology that is produced dialectically. The process of gendering robots makes especially clear that gender belongs both to the order of the material body and to the social and discursive or semiotic systems within which bodies are embedded." [21] In her work on Japanese robotics she finds that while many Japanese robotics initiatives may appear innovative, "it is not new values but rather the renewal of old values - especially those represented by the patriarchal extended family and wartime ideologies - that constitute the significant changes associated with the robotization of society." [22] Through a close reading of details like the pitch and prosody of Japanese fembot voices, Roberts argues that most humanoid robotics researchers and policy makers working under national research initiatives like *Innovation 25* approach humanoid robots from "unprogressive notions of gender dynamics and the sexual division of labor, along with discriminatory attitudes toward non-Japanese migrant workers." [21]

Art robots offer alternatives to robot subjugation by creating diverse roles for machines, and freeing them from subjugation. Art robots largely escape the servile utility that characterize research robots, by being "useless," or ambiguously purposed. Norman White, "the first artist to have consistently championed robotics as an art form," [23] successfully countered robot subjugation in 1985 with his project *Helpless Robot*, which wheedles, cajoles, and commands viewers to help it. "This work humorously reverses the polarity of robot-human relationships, asking humans to help an electronic creature conventionally designed to be a human aid." [ibid] Over the years White has rebuilt the robot in different forms, but in 2012 the "*Helpless Robot* [was] made of steel, wood and [had] handles to move him. There is no motor in the construction, but it has sensors and a synthetic voice that asks you to touch and move it about… The robot says things like: I appreciate your help but you are turning me too far, I said: go to the right! Go back I said, you can turn me now to the left. The personality does not instruct the audience at random, but goes through different phases, from friendliness to grumpiness." [24]

Like Paik's hermaphroditic robot, art robots typically have more gender diversity than mainstream robots, and they can be confusingly sexual. Berlin-based artist Laura Kirkauka invited White to co-develop a sex robot couple in 1988. Each robot were was designed by each artist without coordination, except for standardized genital dimensions. Each artist saw the other's robot for the first time only when they conjoined in a crowded public performance. [25] Unlike the depressingly myriad history of sexualized, subjugated female robots, well-categorized by Wosk [26], Kirkauka's female robot was not designed for the human male gaze, but rather constructed from "a diverse assemblage including a boiling kettle, a squirting oil pump, a twitching sewing machine treadle, and huge solenoid on a fur-covered board — all hanging from an old bedspring and energized by an electronic power sequencer." [27] White, meanwhile, designed a male robot that was more anthropomorphic and literal, with hip motors and an "orgasm strobe." Critically, *them fuckin' robots* are not designed to fulfill human desires, but rather their author's speculation as to what each robot would desire in its mate.

Can HRI avoid the traps that Hampton and Robinson have mapped by conceiving of robots that interact with humans for their own fulfillment, rather than for ours? What would such a robotics look like?

## V. Robot Mortality

If slavery is one fundamental design flaw in contemporary robotics, another is surely novelty. The benefits of technological novelty are often overstated, even while their collateral impacts are credibly criticized by environmentalists, labor activists, privacy advocates, and others. The examples of robot art described above have all reused and repurposed existing materials – often ones with strong cultural significations – rather than seeking the clean, modernist abstraction typical of new industrial products. Though each artist above may describe their rejection of the trappings of novelty differently, and while artists' highly constrained budgets certainly factor into their urge to reuse, it's clear that most artists' professional identity formation includes an appreciation of storied, worn, or even obsolete technology. In contrast to engineering, which relies on contemporary journals and constantly revised textbooks, art education tends to encourage both attention to history and reverence for primary sources. The relative junkiness of many art robots may thus be seen not as a superficial surface consideration, but rather as a manifestation of specific knowledge. Indeed, much as whiteness and maleness are often falsely dissimulated as neutral, laboratory aesthetics are neither neutral, basic, nor inherently technically superior. It may be easy to imagine that structural aluminum extrusions or vacuum-bagged carbon fiber composite have been chosen for their excellent strength-to-weight ratio and robustness, but the laboratory robots these materials are used to build rarely run for more than a few weeks of uptime. Tinguely's art robots built from early 20[th] Century trash have been running on-and-off for seventy years.

The passé, shopworn exteriors of many art robots embed a critique of innovation, and reverberate with what Jackson has described as *broken world thinking*, which foregrounds "erosion, breakdown, and decay, rather than novelty, growth, and progress as our starting point in thinking through the nature, use, and effects of information technology and new media." [28] Indicating both an aesthetic sensibility and a way of knowing, Jackson describes the appreciation and recognition of "subtle arts of repair by which rich and robust lives are sustained against the weight of centrifugal odds, and how sociotechnical forms and infrastructures, large and small, get not only broken but restored..." [ibid] Art robots may be viewed through Jackson's lens as an act of repair to a flawed robotics research agenda.

## VI. Conclusion

Can HRI researchers who seek to develop critical robotics benefit from the lessons that critical roboticists from fine art have learned? There are many institutional, epistemological, and methodological differences between the fields that may attenuate or limit potential benefits. As this paper has indicated, factors that may prevent adoption include that: fine art robots may be "nightmares" for HRI practitioners from engineering backgrounds; while art seeks and attains truths, it does so under very different premises from scientific and technical investigations [29]; the fields that contribute to HRI may not be ready to discard utility, or conduct experiments *for* rather than *on* human subjects.

Nonetheless, critical and speculative approaches have helped to transform HCI into a far more robust and productive discipline, and have come in part from outside what had previously been defined as central to the field.[30][31] While we may not be able to see what successful hybrids and multiplicative effects might be generated by cross-pollination between art robots and HRI, this list of potential avenues to explore below is offered as a friendly gesture to fellow travelers:

- Use found and culturally meaningful objects rather than aluminum extrusions or fresh vacuum-formed plastic
- Read one historical text for every contemporary paper
- Develop a robot for a specific subculture, or a specific location with meaning to the people it will be interacting with
- Consider giving your robot a non-binary gender, a complicated identity, or progressive politics that might upset defense-oriented funders

- Investigate your city's, university's, or family's relationship to servitude and factor that into your research
- Make an emancipated robot that doesn't serve you or other humans; design self-possession and desire into your robot
- Find humans that aren't experimental subjects, and offer them transformation without expecting anything in return
- Set utility aside for a while to see what reveals itself